\documentclass[pra,showpacs,amsmath,a4paper,twocolumn,superscriptaddress]{revtex4}
\usepackage{geometry}
\usepackage{amssymb}
\usepackage{amsmath}
\usepackage{graphicx}
\usepackage{amsfonts}
\usepackage{hyperref}
\usepackage{oldgerm}
\usepackage{color}
\usepackage{epsfig}
\usepackage{bbm}
\usepackage{bbold}

\DeclareMathOperator{\Tr}{Tr}

\begin{document}
\title{The entropy vector formalism and the structure of multidimensional entanglement in multipartite systems}
\author{Marcus Huber}
\affiliation{Fisica Teorica: Informacio i Fenomens Quantics,, Universitat Autonoma de Barcelona, 08193 Bellaterra, Barcelona, Spain}
\affiliation{ICFO-Institut de Ciencies Fotoniques, Mediterranean Technology Park, 08860 Castelldefels (Barcelona), Spain}
\author{Mart\' i Perarnau-Llobet}
\affiliation{ICFO-Institut de Ciencies Fotoniques, Mediterranean Technology Park, 08860 Castelldefels (Barcelona), Spain}
\author{Julio I. de Vicente}
\affiliation{Departamento de Matem\'aticas, Universidad Carlos III de Madrid, Avda.\ de la Universidad 30, 28911, Legan\'es (Madrid), Spain}
\begin{abstract}
We review and generalize the recently introduced framework of entropy vectors for detecting and quantifying genuine multipartite entanglement in high dimensional multicomponent quantum systems. We show that these ideas can be extended to discriminate among other forms of multipartite entanglement. In particular, we develop methods to test whether density matrices are: decomposable, i.\ e.\ separable with respect to certain given partitions of the subsystems; $k$-separable, i.\ e.\ separable with respect to $k$ partitions of the subsystems; $k$-partite entangled, i.e. there is entanglement among a subset of at least $k$ parties. We also discuss how to asses the dimensionality of entanglement in all these cases.
\end{abstract}


\maketitle
\section*{Introduction}
Entanglement is one of the central objects of interest in Quantum Information Theory \cite{nielsen}. Its mere existence implies grave consequences for the foundations of physics, realized already at the dawn of quantum mechanics. Since its first experimental verification decades later it was found that this abstract property of quantum states can be exploited to enable exponential speed up in computing various algorithms \cite{JoszaLinden,MBQC,Shor} and provide the basis for quantum key distribution (QKD) \cite{Ekert91,DIQKD,HP}. Since entanglement has been a topic of focus it has been realized that it also plays a role in critical properties of condensed matter systems and is involved in many other complex physical systems \cite{phase1,helium,AlickiFannes,HPHA,fridge}, possibly even biological ones \cite{bio}.\\
While being an ubiquitous feature of the quantum phase space, entanglement is far from characterized completely, especially in large and complex systems. Of the undeniably rich structure of multipartite quantum correlations \cite{guhne} very little is understood, however many examples, such as measurement based quantum computation \cite{MBQC} or many other algorithms \cite{brussalgo} suggest these properties to be very useful.
In the modern approach of quantum information theory, entanglement is understood as a resource to overcome the limitation of parties restricted to a particular form of manipulations: local operations and classical communication (LOCC). However, while the bipartite case is well understood, the picture is much less clear in the multipartite regime. For instance, for four parties or more there are infinitely many LOCC classes \cite{vdmv,Julionew} and there are only partial connections between those and schemes to exploit their correlations in a useful way. \\
Entropies and entropy inequalities have proven to be a versatile tool in describing the correlations and amount of information contained in an entangled state. In the bipartite case entropy based entanglement measures can even give precise answers to asymptotic rates of entanglement creation by LOCC, in this case the entanglement of formation (EOF) \cite{EOF}. Another useful LOCC monotone in bipartite systems is the Schmidt number \cite{sn}, the minimal dimension of entanglement one needs to create a given quantum state via LOCC. While it is known that entropy based measures can be rather small and still permit exponential speed up in quantum computation \cite{VandenNest} it remains a fact that the Schmidt number is an essential ingredient.\\
A natural generalization of entropy based entanglement measures (and the Schmidt number) in multipartite systems is given by entropy vectors, including the entropies of all possible reductions of multiparticle quantum states. The entries of these vectors satisfy a rich set of constraints, at least for the von Neumann entropy and the Renyi-0-entropy \cite{Evec1,Evec2,Evec3,Evec4}. In a recent paper \cite{HV2} it was shown how to use the notion of entropy vectors to extend the Schmidt number to multipartite states and characterize the full entanglement dimensionality (see also \cite{multisn} for earlier results). Using this, the authors were able to show that there exists a rich LOCC based structure, that is discrete and can be experimentally distinguished in an efficient way.\\
In this paper we generalize the approach of Ref.~\cite{HV2} and show that using different subsets of entropy vectors one can build a framework for entanglement detection, quantification and classification in multipartite quantum systems elucidating the rich structure of partially entangled states. Furthermore we derive a set of witnesses lower bounding these measures in an experimentally feasible way. Some illustrative examples are then solved to show how these witnesses help to decide (i) the $k$-separability of the state and the level of multipartite entanglement, (ii) which particular decompositions the state admits, and (iii) lower bounds on the dimensionality necessary to create the entangled state.\\
Before proceeding any further, let us properly define and clarify these points. A pure state $|\psi\rangle$ is called $k$-separable if it can be written as a tensor product of $k$-vectors, i.e. $|\psi\rangle=|\phi_1\rangle\otimes|\phi_2\rangle\otimes\cdots\otimes|\phi_k\rangle$. This means that a state is not entangled along at most $k$ subsets of parties. A mixed state can be created from $k$-separable pure states and LOCC if its density matrix can be decomposed into a convex combination of $k$-separable pure states. Thus we call such mixed states also $k$-separable even if the pure states in the decomposition are $k$-separable with respect to different subsets of parties. A widely sought after question is whether a mixed state is $1$-separable, i.e. there is not even a decomposition into at least biseparable states, and these states are called genuinely multipartite entangled (GME).\\
Another issue with partial separability in terms of a resource theory is the question of $k$-separability versus genuine $k$-partite entanglement. A pure state $|\psi\rangle=|\phi_1\rangle\otimes|\phi_2\rangle\otimes\cdots$ with a certain degree of separability is said to be $k$-producible if every vector $|\phi_i\rangle$ involves at most $k$ parties. A pure state is then said to be $k$-partite entangled if it is not $(k-1)$-producible. This means that a subset of at least $k$ parties needs to be entangled. The extension to mixed states is straightforward on the analogy of $k$-separable states. While an $n$-partite (mixed) GME state is necessarily $n$-partite entangled, if for example it is found to be biseparable it remains to be resolved what kind of multipartite entanglement is necessary to create this state: being biseparable just implies that this number ranges from $\lceil n/2\rceil$ to $n-1$.\\
Also, besides the level of multipartite entanglement, one might be interested in which particular parties are entangled for a complete characterization of the state. Indeed, consider a tripartite state $\sigma_{\rm ABC}$ which is known to be biseparable. It remains then to know which particular decomposition it admits, for instance, can it be expressed as $ \sigma=p_1 \sigma_A \otimes \sigma_{\rm BC} +p_2 \sigma_{\rm AB} \otimes \sigma_C $? We will call a state decomposable with respect to a set of partitions of the parties if it can be written as a convex combination of states which are separable with respect to any of these (and only these) partitions.\\
Finally, let us comment on the dimensionality of the state. For higher dimensional systems there is no indication of how many dimensions are essentially necessary to create such an entangled state. As an example consider $\frac{1}{\sqrt{2}}(|000\rangle+|111\rangle)$ and $\frac{1}{\sqrt{3}}(|000\rangle+|111\rangle+|201\rangle)$. Both states are genuinely multipartite entangled while the latter requires an extra dimension in subsystem $A$ and is in a certain sense more entangled than the former. This is because the first state can be created by LOCC from the second state but not the other way around. For a precise definition of multipartite entanglement dimensionality see \cite{HV2}.
\\
The questions concerning partial separability have been studied in various contexts before, but this is the first attempt to develop a tool, based on the entropy vector, which allows us to test all of them in a systematic and quantitative way. Good examples of prior approaches are e.g. a set of inequalities to detect both separability and multipartite entanglement of multiqubit states developed in \cite{SeevUff,GuSeev,HMGH,GHH} (see also references therein for an extensive list of partial separability detection criteria). Also in Ref.~\cite{Szi} a relation between subsystem entropies and partial separability was introduced, however a set of accessible bounds was missing. Finally the concept of $k$-producibility and $k$-partite entanglement was introduced in Ref.\cite{seeuff}, and further developed and named in Ref.~\cite{gutobri}, where it was studied in spin chains. General conditions for $k$-partite entanglement have also been recently provided in Ref.\cite{vihyeg} through spin-squeezing inequalities.

\section*{Entropy vectors and lower bounds}
Let us first review the concept of entropy vectors. We will consider states of $n$ parties and $r$ will denote a particular subset of parties, i.\ e.\ a subset of $\{1,2,\ldots,n\}$, and $\overline{r}$ its complement. We will often consider different choices of subsets of parties $r$ that we will group into another set that we will denote by $\mathcal{R}$. For pure states we define the vector $\vec{\mathcal{S}}_\alpha(|\psi\rangle,\mathcal{R})$ using the reductions $\rho_r:=\text{Tr}_{\overline{r}}(|\psi\rangle\langle\psi|)$ with $r\in \mathcal{R}$ via
\begin{align}
\mathcal{S}^r_\alpha(|\psi\rangle,\mathcal{R}):=S_\alpha(\rho_r)
\end{align}
for all $r\in\mathcal{R}$ and where $S_\alpha(\rho):=\frac{1}{1-\alpha}\log_2(\text{Tr}(\rho^\alpha))$. These quantities will give rise to the $|\mathcal{R}|$-dimensional real vector $\vec{\mathcal{S}}_\alpha(|\psi\rangle,\mathcal{R})$, which is ordered such that $\mathcal{S}^j_\alpha\geq \mathcal{S}^{j+1}_\alpha$. For pure states this choice is of course arbitrary, but as we will see this is crucial to capture the multipartite nature of entanglement in mixed states. For these we will, analogously to EOF, take the convex roof over the average ordered entropy vector, i.e.
\begin{equation}\label{snumbervec}
\mathcal{S}^j_\alpha(\rho,\mathcal{R}):=\inf_{\mathcal{D}(\rho)}\sum_i p_i \mathcal{S}^j_\alpha(|\psi_i\rangle,\mathcal{R})\,,
\end{equation}
where $\mathcal{D}(\rho)$ denotes all possible pure ensemble decompositions of $\rho$, i.\ e.\ $\rho=\sum_ip_i|\psi_i\rangle\langle\psi_i|$. It is now crucial to observe that the number of nonzero elements of $\vec{\mathcal{S}}_\alpha(\rho,\mathcal{R})$ can with an appropriately chosen set $\mathcal{R}$ fully reveal the partial separability properties of $\rho$. Also if $\alpha$ is chosen to be zero one can reveal the full dimensionality structure of multipartite mixed states. Unfortunately even in the bipartite case, where there is only one vector element, this convex roof is extremely difficult to compute in general (in fact we know that even the problem of deciding whether this convex roof is nonzero is NP hard). This implies that even given the full density matrix it will generically be impossible to answer these questions in a feasible time. Fortunately this level of precision is not to be desired anyway, since any performed experiment or even physical model of $n$ qu$d$its will not reveal the full density matrix (as a full state tomography scales with $d^{2n}$). It is therefore desirable to rather develop a framework of lower bounds for these measures
that should be as tight as possible while remaining experimentally feasible.\\
In Refs.~\cite{bound} such a framework was devised and shown to provide reliable lower bounds to Renyi-$2$-entropies of entanglement. More recently in Ref.~\cite{HV2} it was shown that this framework can be adapted to directly lower bound elements of the ordered entropy vector, using only single qu$d$it subsystems. We can now show that this construction can readily be adopted to any possible set $\mathcal{R}$ and thus reveal the full range of questions about partial separability. 
First it shall be instructive to remember that for $|\psi\rangle=\sum_\eta c_\eta|\eta\rangle$, with $\eta$ being a multiindex of $n$ entries taking the values $0$ and $d-1$ we can write the linear 2-entropy ($S_L(\rho):=\sqrt{2(1-\text{Tr}(\rho^2))}$) of the reductions as
\begin{align}
S_L(\rho_{r})^2=\sum_{\eta,\eta'}|c_{\eta}c_{\eta'}-c_{\eta_{r}}c_{\eta'_{r}}|^2
\end{align}
where the pair $(c_{\eta_{r}},c_{\eta'_{r}})$ is just equal to the pair $(c_{\eta},c_{\eta'})$, but with all components of $\eta$ and $\eta'$ that are part of the reduction $r$ exchanged. Using that $|C|\sum_{i\in C}|a_i|^2\geq|\sum_{i\in C}a_i|^2$ \cite{bound} and that $|a-b|\geq|a|-|b|$, we arrive at
\begin{equation}
S_L(\rho_{r})\geq \frac{1}{\sqrt{|C|}} \sum_{\eta,\eta'\in C}(|c_{\eta}c_{\eta'}|-|c_{\eta_{r}}c_{\eta'_{r}}|)
\end{equation}
for any subset $C$ of multiindices of $n$ entries. Therefore, we can bound the ordered linear entropy vector $\mathcal{S}^j_L$ for pure states as
\begin{align}
\mathcal{S}^j_L(|\psi\rangle)\geq \frac{1}{\sqrt{C}} \sum_{\eta,\eta'\in C} (|c_{\eta}c_{\eta'}|-\min_{\{r_m\}\subseteq\mathcal{R}}\sum_{m=1}^{j}|c_{\eta_{r_m}}c_{\eta_{r_m}'}|).
\end{align}
Now we can extend this to mixed states via the observation that $\inf(A-B)\geq\inf A-\sup B$. First, it is clear that
\begin{align}
\inf_{\mathcal{D(\rho)}}\sum_i p_i|c^i_{\eta}c^i_{\eta'}|\geq|\sum_i p_ic^i_{\eta}{c^i_{\eta'}}^*|=|\langle\eta|\rho|\eta'\rangle|.
\end{align}
For the supremum we can use
\begin{align}
\sup_{\mathcal{D(\rho)}}&\sum_ip_i\min_{\{r_m\}\subseteq\mathcal{R}}\sum_{m=1}^{j}|c^i_{\eta_{r_m}}c^i_{\eta_{r_m}'}|\nonumber\\
&\leq\min_{\{r_m\}\subseteq\mathcal{R}}\sup_{\mathcal{D(\rho)}}\sum_ip_i\sum_{m=1}^{j}|c^i_{\eta_{r_m}}c^i_{\eta_{r_m}'}|\nonumber\\
&\leq \min_{\{r_m\}\subseteq\mathcal{R}}\sum_{m=1}^{j}\sqrt{(\sum_ip_i|c^i_{\eta_{r_m}}|^2)(\sum_ip_i|c^i_{\eta_{r_m}'}|^2)}\nonumber\\
&=\min_{\{r_m\}\subseteq\mathcal{R}}\sum_{m=1}^{j}\sqrt{\langle\eta_{r_m}|\rho|\eta_{r_m}\rangle\langle\eta'_{r_m}|\rho|\eta'_{r_m}\rangle}.
\end{align}
In conclusion, we end up with $\mathcal{S}_2^j(\rho)\geq -\log_2(1-\frac{W_j(\rho,C,\mathcal{R})^2}{2})$, where
\begin{align}
W_j(\rho,C,\mathcal{R}):=\frac{1}{\sqrt{|C|}}\sum_{\eta,\eta'\in C}\big[|\langle\eta|\rho|\eta'\rangle|\nonumber\\-\min_{\{r_m\}\subseteq\mathcal{R}}\sum_{m=1}^{j}\sqrt{\langle\eta_{r_m}|\rho|\eta_{r_m}\rangle\langle\eta'_{r_m}|\rho|\eta'_{r_m}}\rangle\big].\label{bounds}
\end{align}
with $j\in \{1,...,|\mathcal{R}| \}$. Since $\mathcal{R}$ is fully determined by the question we ask there are only two freedoms in designing these witness lower bounds. First of all of course the local unitary representation of the state, that should be chosen such that there as few as possible off-diagonal elements in the density matrix. The second freedom concerns the choice of the set of off-diagonal elements $C$. This choice is further constrained by some obvious facts: It should not be invariant under transposing any subsystem that is in the set $\mathcal{R}$, because otherwise it would just add a strictly negative term to the sum in Eq.\ (\ref{bounds}). So e.\ g.\ for $j=1$ and $\mathcal{R}$ being the full set of subsystems only anti-diagonal (orthogonal in every subsystem) elements of the density matrix are a suitable choice for (\ref{bounds}).\\

\section*{Classification of multipartite entanglement}
In this section we provide some examples to show how such bounds would be constructed in practice and the power of their applicability.

\emph{Subsets of the entropy vector and partial decomposability.} Consider a state $\rho$ and a set  $\mathcal{R}=\{r_i\}$ and the corresponding entropy vector (\ref{snumbervec}). Notice that if all entries of (\ref{snumbervec}) are nonzero, then the state cannot be expressed as
\begin{equation}
 \sum_{i} p_i \rho_{r_i} \otimes \rho_{\bar{r}_i},
\label{eq:decom}
\end{equation}
where $\rho_{r_i} \otimes \rho_{\bar{r}_i}$ is separable with respect to the bipartition $(r_i|\bar{r}_i)$. Conversely, $k$ zeroes imply the state being decomposable in states which are at least decomposable in $k$ bipartitions of the set $\mathcal{R}$.
Since the witnesses (\ref{bounds}) provide lower bounds on the entropy vector, we can only ascertain when a state is \emph{not} decomposable
in the form (\ref{eq:decom}). On the other hand, we are free to choose any set $\mathcal{R}$, allowing us to obtain
a full detailed description of the partial decomposability regions of a multipartite entangled state.

\begin{figure*}[htp]

\begin{minipage}[b]{0.48\linewidth}
\centering
\includegraphics[width=\textwidth]{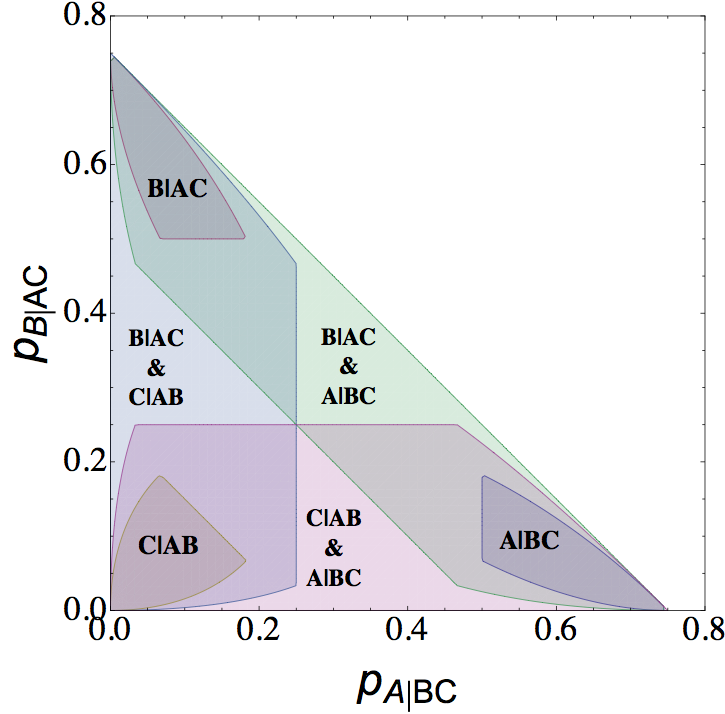}
\end{minipage}
\begin{minipage}[b]{0.49\linewidth}
\centering
\includegraphics[width=\textwidth]{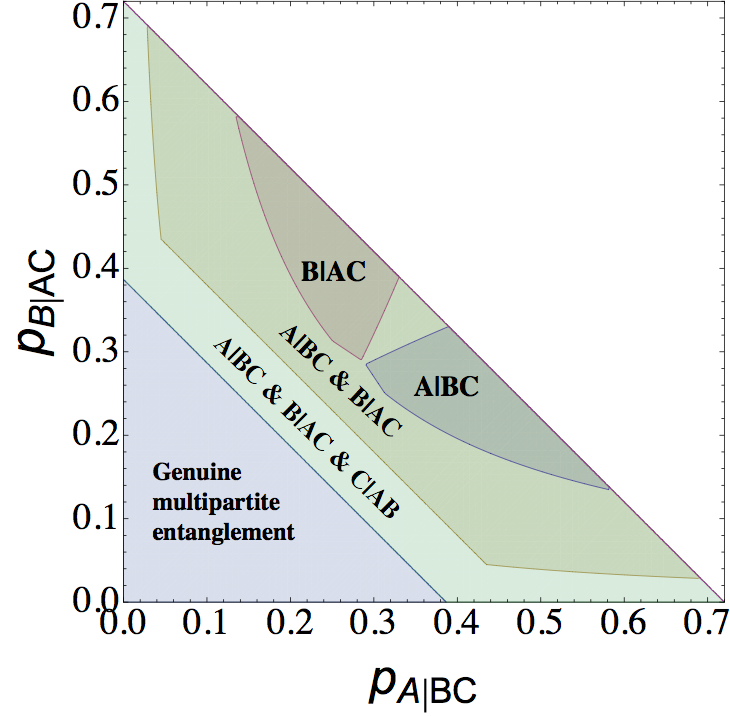}
\end{minipage}
\caption{(Color online) Both figures show the different regions of non-decomposability of the state (\ref{state.exam.1}) using (\ref{bounds}). They can be understood as follows: for instance, every point outside the region $A|BC$ cannot be decomposed as a convex combination of $\sigma^i_A \otimes \sigma^i_{BC}$. In the first figure (left) we set $p_{ABC}=0.25$, so the noise is strong enough to prevent multipartite entanglement. In the second figure (right) we set $p_{C|AB}=0.28$ so that the amount of multipartite entanglement is not fixed. In this figure two new regions appear: a region where the state is multipartite entangled and a region where the state is biseparable but it can only be decomposed in the form $p_1 \sum_i p_i\sigma^i_A \otimes\sigma^i_{BC}+p_2 \sum_i q_i\sigma^i_B \otimes\sigma^i_{AC}+p_3\sum_i r_i \sigma^i_C \otimes\sigma^i_{AB}$. For both pictures we choose the set $C$ such that the witnesses at every point reach maximal value.}
\label{exdecom}
\end{figure*}

Let us apply this idea to the normalized tripartite state of qubits:
\begin{eqnarray}
 \rho_1=p_{A|BC} \frac{ \mathbb{1}_A}{2} \sigma_{BC} + p_{B|AC} \frac{ \mathbb{1}_B}{2}\sigma_{AC}+\nonumber\\
+p_{C|AB} \frac{ \mathbb{1}_C}{2} \sigma_{AB}+ p_{ABC}  \sigma_{ABC}
\label{state.exam.1}
\end{eqnarray}
 where  $\sigma_{AB}=|\phi_{AB}\rangle \langle \phi_{AB} |$, with $|\phi_{AB} \rangle =\frac{1}{\sqrt{2}}(|0_A 0_B\rangle + | 1_A 1_B \rangle)$ and similarly for $\sigma_{BC}, \sigma_{AC}$; and finally
$\sigma_{ABC}=|\phi_{ABC}\rangle \langle \phi_{ABC} |$, with $|\phi_{ABC} \rangle=\frac{1}{\sqrt{2}}(|0_A 0_B 0_C\rangle + | 1_A 1_B 1_C\rangle)$.
Whereas the last term in (\ref{state.exam.1}) contains a pure tripartite entangled state, the remaining terms contain
bipartite entangled states together with a fully mixed state (noise) in the other party.
Now, for example, consider the set $\mathcal{R}_1=\{A,B\}$, we obtain
\begin{eqnarray}
 \mathcal{W}_2(\rho_1,C_1,\mathcal{R}_1)=\frac{1}{\sqrt{3}} \left(p_{C|AB}- 2 \sqrt{p_{A|BC}p_{B|AC}} + \right. \nonumber\\
\left.
   +p_{ABC} - \frac{p_{A|BC}+p_{B|AC}}{2}\right)\label{w.exam.1}
\end{eqnarray}
where we used the set of off-diagonal entries $C_1=\{(000,111),(000,110),(001,111)\}$ in order to maximize the bound.
If ${W}_2(\rho_1,C_1,\mathcal{R}_1)>0$, then so it is the lowest entry of the entropy vector in the subset $\mathcal{R}_1$;
and therefore the state cannot be decomposed as $ \sigma=p_1 \sigma_A \otimes \sigma_{\rm BC} +p_2 \sigma_{\rm AC} \otimes \sigma_B $.
Similar conditions can be obtained for other choices of $\mathcal{R}$ in order to elucidate the different
partial decomposability regions, as illustrated in figure \ref{exdecom}.

To our knowledge, these are the first tools aimed at identifying partial decomposability in a systematic way. An application of this is given by the problem of deciding when a channel is entanglement annihilating (EA). A channel $\Phi^{AB}$ acting on bipartite input states $\rho_{AB}$ is EA if its output $\Phi^{AB}(\rho_{AB})$ is disentangled for all inputs \cite{ea1}. Let $\Omega^{ABA'B'}_\Phi$ be the 4-partite Choi-Jamiolkowski (CJ) state associated to the channel $\Phi$, i.\ e.\ $\Omega^{ABA'B'}_\Phi=\Phi^{AB}\otimes\mathbb{1}^{A'B'}(|\phi^+\rangle_{AA'}|\phi^+\rangle_{BB'})$ where $|\phi^+\rangle=(|00\rangle+|11\rangle+\cdots+|d-1,d-1\rangle)/\sqrt{d}$ is the 2--qudit maximally entangled state. It has been recently shown \cite{ea2} that if $\Omega^{ABA'B'}_\Phi$ is decomposable in the bipartitions $A|BA'B'$ and $B|AA'B'$, then the channel is EA. Unfortunately, in principle our techniques do not allow us to conclude whether a channel is EA or not because they allow us to decide when a state is not decomposable while decomposability is a sufficient condition for EA. However, if following \cite{ea2} we apply our tools to the 2--qubit local depolarizing channel $\Phi^A_{q_1}\otimes\Phi^B_{q_2}$ and the 2--qubit global depolarizing channel $\Phi^{AB}_{q}$ for which
\begin{align}
\Omega&_{\Phi^A_{q_1}\otimes\Phi^B_{q_2}}=q_1q_2|\phi^+\rangle_{AA'}\langle\phi^+|\otimes|\phi^+\rangle_{BB'}\langle\phi^+|\nonumber\\
&+q_1(1-q_2)|\phi^+\rangle_{AA'}\langle\phi^+|\mathbb{1}_{BB'}/4\nonumber\\
&+(1-q_1)q_2\mathbb{1}_{AA'}|\phi^+\rangle_{BB'}\langle\phi^+|\nonumber\\
&+(1-q_1)(1-q_2)\mathbb{1}_{AA'BB'},\nonumber\\
\Omega&_{\Phi^{AB}_{q}}=q|\phi^+\rangle_{AA'}\langle\phi^+|\otimes|\phi^+\rangle_{BB'}\langle\phi^+|\nonumber\\
&+(1-q)\mathbb{1}_{AA'BB'},
\end{align}
we readily find that
\begin{align}
W_2(\Omega_{\Phi^A_{q_1}\otimes\Phi^B_{q_2}},C_2,\mathcal{R}_2)&=\frac{3q_1q_2-1}{8},\nonumber\\
W_2(\Omega_{\Phi^{AB}_{q}},C_2,\mathcal{R}_2)&=\frac{3q-1}{8},
\end{align}
where of course $\mathcal{R}_2=\{A,B\}$ and $C_2=\{(0000,1111)\}$. This implies that both CJ states are not decomposable in the splittings $A|BA'B'$ and $B|AA'B'$ when $q_1q_2>1/3$ and $q>1/3$ respectively. Interestingly this is precisely the threshold in which the channels become EA \cite{ea2,ea3}. This suggests that our condition might be necessary and sufficient for these states and encourages further research in this direction.

\emph{Separability}.
Sometimes one is not interested in which specific decomposition a state admits, but rather on the partial $k$-separability of the
state (i.e., whether we can find a decomposition with states that are at most $k$-separable). It is then enough to consider
the \emph{full} entropy vector made up of $2^{N-1}-1$ elements (i.e., the total number of bipartitions). Then, it is easy to
realize that a $k$-separable state will have $2^{k-1}-1$ zero entries in the full entropy vector. For our bounds this implies that
$\rho$ is at most $k$-separable -or, equivalently, $\rho$ is not $k+1$-separable- if
\begin{equation}
\label{condsep}
 W_{\gamma(k)}(\rho,C,\mathcal{R}_{\rm total})>0, \hspace{7mm} \gamma(k)=2^{N-1}-2^k+1, 
\end{equation}
where $k\in \{1,2,..,N-1 \}$ and $\mathcal{R}_{\rm total}$ corresponds to the set of all bipartitions.

\begin{figure*}[ht]
\begin{minipage}[b]{0.49\linewidth}
\centering
\includegraphics[width=\textwidth]{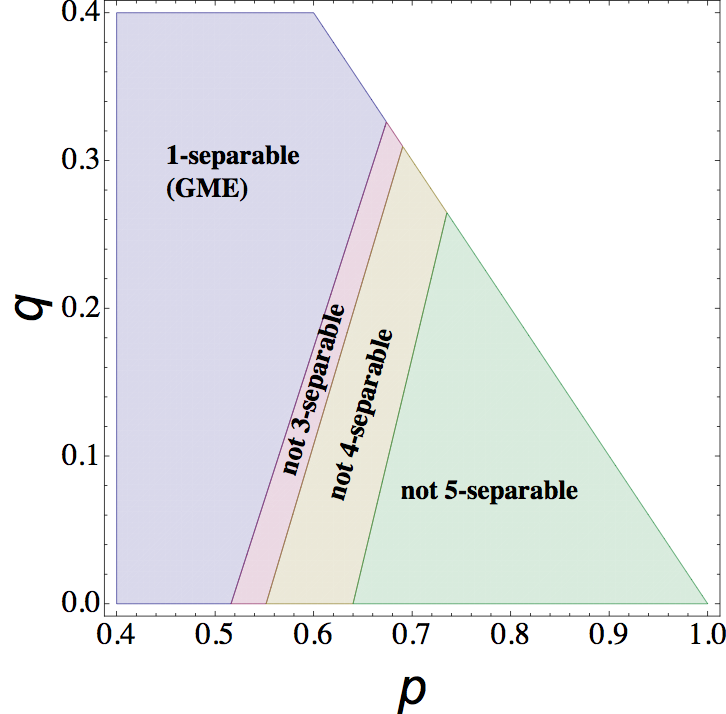}
\end{minipage}
\begin{minipage}[b]{0.49\linewidth}
\centering
\includegraphics[width=\textwidth]{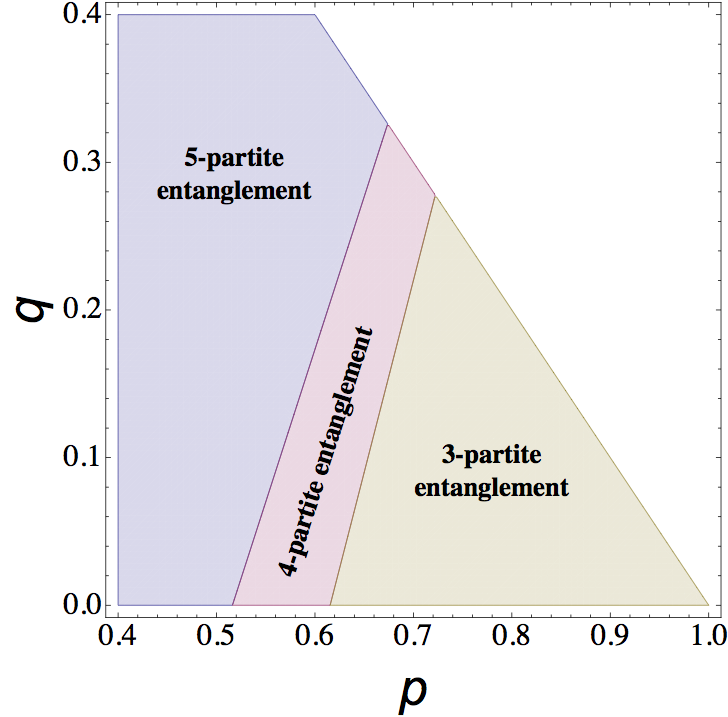}
\end{minipage}
\caption{(Color online) The left (right) plot shows the different regions of $k$-separability (multipartite entanglement) as derived by our witnesses for (\ref{examstateII})with $N=5$ and $\alpha=\beta=1/\sqrt{2}$. Notice the left figure show regions of non-separability, so that, for instance, not $4-$separable is equivalent to 1, 2 or 3-separable. On the other hand, in the right figure we show regions where at least $k$-multipartite entanglement is present. }
\label{fig:multpart}
\end{figure*}

As an application let us consider the $N$-partite state:
\begin{equation}
 \rho_2 = (1-p) \rho_{\rm GHZ}+ q \rho_{\rm dep}+ (p-q)  \frac{\mathbb{1}}{2^N}
\label{examstateII}
\end{equation}
where $\rho_{\rm GHZ}=|{\rm GHZ} \rangle_N \langle {\rm GHZ} |$ with
$|GHZ \rangle_N = \alpha |0_1 0_2 ... 0_N \rangle + \beta |1_1 1_2 ... 1_N \rangle$,
 $\rho_{\rm dep}$ is the dephased version of $\rho_{\rm GHZ}$, and the last term represents white noise.
Expression (\ref{condsep}) is easily computed yielding
\begin{equation}
 W_{\gamma(k)}=2\left((1-p)\alpha \beta -(2^{N-1}-2^{k}+1)\frac{p-q}{2^N}\right),
\end{equation}
which allows us to find restrictions on the regions of partial separability of
$\rho_2$ as illustrated in figure \ref{fig:multpart}.

\emph{Multipartite entanglement}.
Even if we know the regions of partial $k$-separability, an open question is its relation to the presence of genuinely multipartite entanglement.
Indeed, a $N-$partite state which is found to be $k$-separable can exhibit anything from $\lceil N/k \rceil$ to $(N-k+1)$-partite entanglement. The derived witnesses
allows us to set restrictions on the regions of $(\lceil N/2\rceil+m)$-partite entanglement, with $m\in\{0,1,...,\lfloor{N/2}\rfloor-1\}$, using the following procedure: first
one computes the last entry of $W_j(\rho,\mathcal{R}_{N/2})$ where $\mathcal{R}_{h}$ contains all subsets $r$ of exactly $h$ parties. If all elements are nonzero, then one repeats the computation for the set $\{\mathcal{R}_{N/2},\mathcal{R}_{N/2+1}\}$.
This process is iterated until we find a set $\mathcal{G}_{m}=\{\mathcal{R}_{N/2},\mathcal{R}_{N/2+1},...,\mathcal{R}_{N/2+m}\}$ where
 $W_{|\mathcal{G}_m|}$ is zero. This implies that the state contains at least $(\lceil N/2\rceil+m)$-partite entanglement, or equivalently, that it is not $p$-producible with $p =\lceil N/2\rceil+m-1$. Finally, let us note that the size of the subsets
is given by:
\begin{align}
 |\mathcal{G}_m|=
\begin{cases}
\frac{1}{2} \binom {N} {N/2}+\sum_{i=1}^{m} \binom{N}{\frac{N}{2}+i}, & \text{if }N\text{ is even} \\
\sum_{i=0}^{m} \binom{N}{\frac{N+1}{2}+i}, & \text{if }N\text{ is odd}
\end{cases}
\end{align}
Finally, notice that if no zero is found in $W_{|\mathcal{G}_m|}$ then the state is $N$-partite entangled. This procedure is applied to the state (\ref{examstateII}), as we can see in figure (\ref{fig:multpart}).

\emph{Detecting higher dimensionality}. Finally we want to point out the connection to the entanglement dimensionality vector introduced in Ref.~\cite{HV2}. Since $\mathcal{\vec{S}}_\alpha$ is a non-increasing function of $\alpha$ in every element, Eq.\ (\ref{bounds}) directly provides a lower bound on the dimensionality vector $\mathcal{\vec{S}}_0$ as well. Of course, this is just a lower bound and it can be rather crude when the distribution of eigenvalues is not particularly flat. However, this is to be expected from a witness, as in the space of density matrices all dimensionalities lie within an $\epsilon$-small region. Distinguishing those is not the goal of any experimentally feasible witness as this would lie beyond any error margin of actual measurement outcomes. Furthermore for accessing the full dimensionality in e.g. cryptography, where it is desirable that a single measurement on a $d$-dimensional state leads to $\log_2(d)$ random bits it is actually necessary to have flat distributions at hand.

\begin{figure}[ht!]
  \centering
    \includegraphics[width=0.48\textwidth]{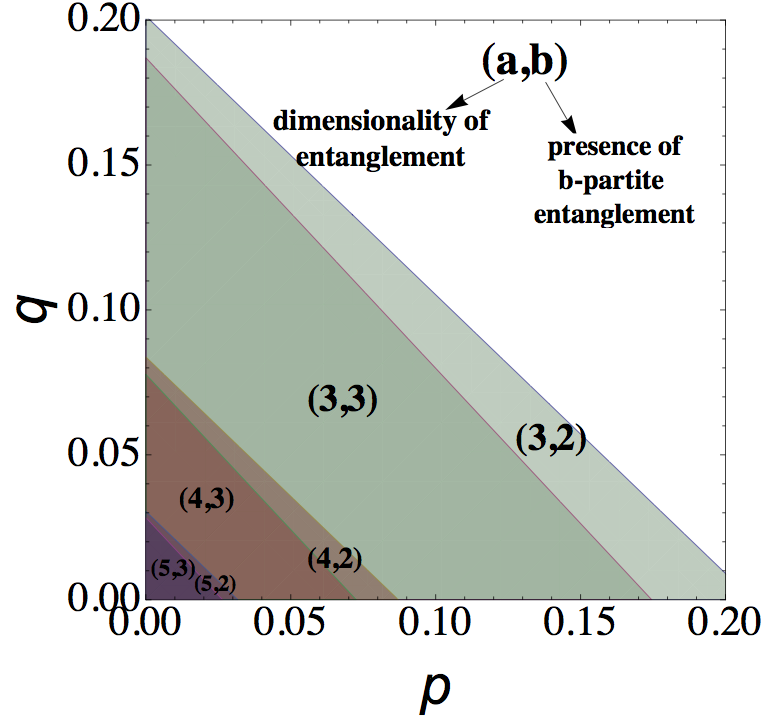}
    \caption{(Color online)Regions of partial decomposability and effective dimensionality of entanglement for the state $\rho_3$ defined in (\ref{rho3}).}
\label{fig:dim}
\end{figure}

A suitable examplary case is given by the following five party qutrit state:
\begin{equation}
	\rho_3=(1-p-q) \rho_{\rm GHZ}^{(5,3)} +p\frac{ \mathbb{1}_A}{5}\rho_{\rm GHZ}^{(5,2)} +q\frac{ \mathbb{1}_{\rm ABC}}{5^3}
	\label{rho3}
\end{equation}
where $\rho_{\rm GHZ}^{(a,b)} $ stands for a GHZ state of $b$ parties with local dimension $a$. We want to illustrate how our framework can be used to reveal genuine multidimensional $m$-partite entanglement in a multipartite system. We depict the details in figure \ref{fig:dim}.

\section*{Enhancing the detection by applying local invertible operators}

It is straightforward to notice that all partial separability/entanglement properties considered here remain invariant under local transformations of the form
\begin{equation}\label{fnf}
\rho\rightarrow\frac{A_1\otimes A_2\otimes ... \otimes A_n\rho A_1^\dag\otimes A_2^\dag\otimes ... \otimes A_n^\dag}{\Tr(A_1\otimes A_2\otimes ... \otimes  A_n\rho A_1^\dag\otimes A_2^\dag\otimes  ... \otimes A_n^\dag)}
\end{equation}
if the local operators $\{A_i\}$ are all invertible. This holds also true for the entanglement dimensionality because such mapping cannot change the local ranks of every pure state in each ensemble decomposition of $\rho$. This idea has already been used in the context of the separability problem in \cite{Git} to increase the detection of several bipartite entanglement tests and in \cite{jens} to compute a three-qubit entanglement measure. The idea is to choose the local operators such that the state is mapped to the normal form of \cite{frank} in which all single party reductions are proportional to the identity (or arbitrarily close to it). The intuition is that entanglement in the normal form (whenever it exists) is in a certain sense maximized and more likely to be detected by the different criteria. Moreover, given the state, the local operators mapping it to the normal form can be efficiently computed \cite{frank}. However, this of course requires full knowledge of $\rho$ and it is therefore of limited use in practical situations where tomography is unaffordable. Nevertheless, from a purely theoretical point of view, one may wish to determine very precisely the limits for which certain families of states have certain separability or entanglement properties. As argued above, the transformation (\ref{fnf}) can also be used in our context to improve our conditions. Furthermore, this can also be used to asses the performance of the original inequalities.

To see a simple example of this, consider the (for convienience unnormalized) state
\begin{align}
\sigma&=p_{A|BC}M_A\chi_{BC}+p_{B|AC}M_B\chi_{AC}+p_{C|AB}M_C\chi_{AB}\nonumber\\
&+p_{ABC}\omega_{ABC},\label{statef}
\end{align}
where $\chi$ ($\omega$) is the density matrix associated to the pure state $\sqrt{2}|00\rangle+\sqrt{2}/4|11\rangle$ ($2|000\rangle+1/4|111\rangle$) and $M=diag(1,1/4)$. The local operators $A_1=A_2=A_3=diag(\sqrt{2},1/\sqrt{2})$ map $\sigma$ to its normal form, which is given by $\rho_1$ in Eq.\ (\ref{state.exam.1}). Hence, although using the same witness as in that case to check decomposability in $A|BC$ and $B|AC$ we obtain
\begin{equation}\label{w.exam.1f}
p_{ABC}+\frac{5}{4}p_{C|AB}-\frac{1}{2}(p_{A|BC}+p_{B_AC})-\frac{5}{2}\sqrt{p_{A|BC}p_{B|AC}}>0,
\end{equation}
we know that $\sigma$ is not $\{A|BC,B|AC\}$-decomposable whenever $\rho_1$ is not. Thus, in general, if we want to decide whether $\sigma$ has this property, it is preferable to map to the normal form $\rho_1$ and use Eq.\ (\ref{w.exam.1}) instead of (\ref{w.exam.1f}). However, arguing in the opposite direction, it is interesting to notice that for very small values of $p_{A|BC}$ or $p_{B|AC}$, condition (\ref{w.exam.1f}) is more powerful than (\ref{w.exam.1}), showing us that our conditions for $\rho_1$ could still be improved in these ranges.

The mapping (\ref{fnf}) can also be very useful to detect entanglement dimensionality. Recall from our discussion above that this is particularly difficult when the distribution of local eigenvalues is very peaked. In this case the local operators can be used to flatten this distribution. To illustrate what we mean consider the 3-qutrit state
\begin{equation}
|\psi_\epsilon\rangle=\sqrt{1-2\epsilon}|000\rangle+\sqrt{\epsilon}|111\rangle+\sqrt{\epsilon}|222\rangle,
\end{equation}
where $\epsilon$ is small. Nevertheless, the entanglement dimensionality vector for this state is by construction $(3,3,3)$ as the state is genuinely entangled in all three dimensions. However, if we take for instance $\epsilon=0.1$ our conditions yield $\vec{S_L}\geq(0.8,0.8,0.8)$ and we can only conclude that the state contains $(2,2,2)$ entanglement. On the other hand, applying the local operator $A_1=diag(1/\sqrt{3(1-2\epsilon)},1/\sqrt{3\epsilon},1/\sqrt{3\epsilon})$ the state is mapped to its normal form, the GHZ-like state $(|000\rangle+|111\rangle+|222\rangle)/\sqrt{3}$ for which we obtain $\vec{S_L}\geq2/\sqrt{3}(1,1,1)$ and entanglement dimensionality $(3,3,3)$ is inferred $\forall\epsilon>0$. Of course this example is trivial because we are dealing with pure states, but if we consider now some noisy form $p|\psi_\epsilon\rangle\langle\psi_\epsilon|+(1-p)\mathbb{1}/27$, by mapping to the normal formal we can nontrivially conclude full $(3,3,3)$ entanglement for sufficiently large values of $p<1$.

\section*{Conclusions}
Starting from the framework introduced in Ref.~\cite{HV2} we have derived a general tool to investigate partial separabilities of multipartite systems in arbitrary dimensions. The introduced framework is capable of revealing genuine multipartite dimensionalities of multipartite entanglement and can reveal arbitrary non-decomposabilites of density matrices. The framework is based on identifying significant coherences present in the system and requires at most a square root of the measurements required for a full state tomography. If, however, one approaches the problem from the purely theoretical point of view and assumes full knowledge of the density matrix, we have also shown how our conditions can be improved by suitably transforming the state. We have presented illustrative examples that instructively show how the criterion can be applied and to what extent it reveals partial separabilities and dimensionalities in highly mixed multipartite and multidimensional systems.\\
These results relate in a natural way to the marginal problem of multipartite mixed states \cite{marginalp} and the characterization of multipartite entanglement from properties of the marginals \cite{polytop}, as we give coarse grained answers about possible marginal spectra in terms of their entropy.\\
A natural further line of investigation concerns the possibility of approximating the partially entangled sets and multipartite dimensionalities via semidefinite programming approaches as e.g. \cite{taming} and in such a way possibly extending them to a device independent setting as in \cite{dieq}.\\
\noindent\emph{Acknowledgements.}
MH acknowledges funding from the MC grant "Quacocos". J.I. de V. acknowledges partial support from the Spanish MINECO through grants MTM 2010-21186-C02-02 and MTM2011-26912. M.P.L. acknowledges financial support from the Severo Ochoa program.

\end{document}